\documentclass[10pt,showpacs,notitlepage,superscriptaddress,prd,aps,nofootinbib,longbibliography]{revtex4-1}
\usepackage{graphicx}\usepackage{amsmath,amssymb,dsfont}\usepackage{slashed}
\usepackage{epsfig}
\usepackage{epstopdf}
\usepackage{adjustbox}
\usepackage{color}
\definecolor{darkgreen}{rgb}{0,.5,0}
\usepackage[colorlinks,filecolor=blue,citecolor=darkgreen,unicode]{hyperref}

\newcommand{\ii}{\mathrm{i}}
\newcommand{\tga}{\tilde\gamma}
\newcommand{\sD}{\slashed D}
\newcommand{\mfm}{\mathfrak{m}}
\newcommand{\Gh}{H}

\begin{document}
\title{Quantum Dirac fermions in half-space \\ and their interaction with electromagnetic field}
\affiliation{Physics Department, Ariel University, Ariel 40700, Israel}
\affiliation{CMCC-Universidade Federal do ABC, Santo Andr\'e, S.P., Brazil}
\affiliation{Dipartimento di Fisica ``Ettore Pancini'', Universit\`{a} di Napoli {\sl Federico II}, 
Napoli, Complesso Univ. Monte S. Angelo, Via Cintia, I-80126 Napoli, Italy}
\affiliation{INFN, Sezione di Napoli, Via Cintia, 80126 Napoli, Italy}
\affiliation{Department of Physics, Tomsk State University, 634050 Tomsk, Russia}
\affiliation{Physics Department, Ariel University, Ariel 40700, Israel}

\author{I. Fialkovsky\footnote{On leave of absence from CMCC-Universidade  Federal  do  ABC,  Santo  Andre,  S.P.,  Brazil}}
\email{ifialk@gmail.com}
\affiliation{Physics Department, Ariel University, Ariel 40700, Israel}

\author{M. Kurkov}
\email{max.kurkov@gmail.com, corresponding author}
\affiliation{Dipartimento di Fisica ``Ettore Pancini'', Universit\`{a} di Napoli {\sl Federico II}, 
Napoli, Complesso Univ. Monte S. Angelo, Via Cintia, I-80126 Napoli, Italy}
\affiliation{INFN, Sezione di Napoli, Via Cintia, 80126 Napoli, Italy}

\author{D. Vassilevich}
\email{dvassil@gmail.com}
\affiliation{CMCC-Universidade Federal do ABC, Santo Andr\'e, S.P., Brazil}
\affiliation{Department of Physics, Tomsk State University, 634050 Tomsk, Russia}

\begin{abstract}
We study the polarization tensor of a Dirac field in $(3+1)$ dimensions confined to a half-space -- a problem motivated by applications to the condensed matter physics, and to Topological Insulators in particular. Although the Pauli-Villars regularization scheme has a number of advantages, like explicit gauge invariance and decoupling of heavy modes, it is not applicable on manifolds with boundaries. Here, we modify this scheme by giving an axial mass to the regulators and to the physical field. We compute the renormalized polarization tensor  {in coordinate representation}. We discuss then the induced Chern-Simons type action on the boundary and compare it to the effective action of a $(2+1)$ dimensional Dirac fermion.
\end{abstract}
%\pacs{...}
\maketitle

\section{Introduction}\label{sec:intro}
Various applications to the physics of new materials sparkled a lot of interest to Quantum Field Theory (QFT) with boundaries or interfaces. A recent example is the mixed dimensional QED \cite{Gorbar:2001qt,Hsiao:2017lch,Vozmediano:2010fz,Kotikov:2013eha,Herzog:2018lqz} in which the photons propagate in the $(3+1)$-dimensional Minkowski space while the fermions are confined to a $(2+1)$-dimensional surface. This model describes graphene interacting with the usual Maxwell field. QFT computations \cite{Bordag:2009fz,Fialkovsky:2011pu} of the Casimir interaction of graphene may be considered as a resummation of certain types of Feynman diagrams in this model. Such QFT computations are in a very good agreement \cite{Klimchitskaya:2014axa} with the experiment \cite{Banishev:2013}, which demonstrates once again the efficiency of QFT in describing the physics of advanced materials.

The model, that we consider in this work describes the Dirac fermions confined to a half-space in $(3+1)$ dimensions interacting with the photons propagating in the whole space. This is a field theory model of Topological Insulators \cite{Tkachov2013}. Only the fermions will be quantized. We shall concentrate on a single Feynman diagram that gives the polarization tensor of external electromagnetic field. This quantity is definitely of a practical interest since it describes the conductivity of Topological Insulators. Besides, there are some more theoretical questions to be answered. One of them is related to the induced Chern-Simons action on the boundary leading to a Hall type conductivity. In some range of the parameters, Dirac fermions on $(3+1)$-dimensional manifolds with boundaries have {surface states}, that are $(2+1)$-dimensional fermionic modes. The fermions in $(2+1)$ dimensions possess a parity anomaly \cite{Niemi:1983rq,Redlich:1983dv,AlvarezGaume:1984nf} that leads to the Chern-Simons action of level $k=1/2$ for the electromagnetic field. Some authors \cite{Mulligan:2013he} provided arguments that the $(2+1)$ dimensional parity anomaly does not lead to the Chern-Simons action on a boundary in $(3+1)$ dimensions, though their computations used in fact domain walls rather than boundaries. Direct evaluation of the parity anomaly in 4 (Euclidean) dimensions for massless Dirac fermions confirmed the existence of the Chern-Simons term on the boundary for both electromagnetic \cite{Kurkov:2017cdz} and gravitational \cite{Kurkov:2018pjw} fields. Remarkably, the level of the Chern-Simons action appeared to be $1/4$, i.e. a half of the parity anomaly in 3 dimensions. In the present work we are going to resolve this problem by computing the induced boundary Chern-Simons type action for a Dirac fermion having both bulk and boundary mass gaps. 

Most of this paper will actually be dedicated to the development of QFT methods with boundaries. First of all, we shall propose a suitable modification of the Pauli-Villars (PV) regularization scheme. This scheme has many advantages. It preserves the manifest gauge invariance. Besides, the PV subtraction ensures decoupling of massive modes, which is a very reasonable requirement in effective theories and condensed matter applications. However, in the presence of a boundary the PV scheme is not immediately applicable since the usual bulk mass fails to provide a gap to specific boundary excitations. To solve this problem it was suggested \cite{Beneventano:2018tle} to add an axial mass term. In the present work we follow the same approach. As we show below, the axial mass indeed gives a mass gap to all states. Apart from enabling us to use the PV scheme, this mass also allows to describe topological insulators with gapped surface states, see e.g. \cite{Cho:2017sic}. We shall formulate the PV scheme, prove the finiteness of regularized effective action and obtain the renormalized expressions. Although these expressions will appear to be rather long and complicated, we shall be able to extract some simple and interesting physical information from the parity-odd part of the polarization tensor. After integration over the normal coordinates, this tensor will give a Hall type conductivity near the boundary. We shall compare this integrated tensor with the parity-odd part of the polarization tensor for a Dirac fermion in $(2+1)$ dimensions. 

Under a different name, the polarization tensor of electromagnetic field in the presence of boundaries was considered in the condensed matter literature, see e.g. \cite{Newns1970,Marusic2006}. In these papers, however, non-relativistic (non-Dirac) dispersion relations for quasiparticles were used. More recently, for a Dirac field in half-space in $(2+1)$ dimensions a one-point function \cite{Dyakonov:2017zak} and the polarization tensor \cite{Fosco:2018nio} were computed.
A Weyl anomaly induced current on the boundary was studied in \cite{Chu:2018ksb}.

Throughout this work we use the natural units $\hbar=c=1$. To facilitate applications to the condensed matter problems we introduce the Fermi velocity $v_F$. Since the $v_F$ dependence of polarization tensor may be recovered by using some simple rules, in most of the paper we keep $v_F=1$, but restore $v_F\neq 1$ whenever necessary.
This paper is organized as follows. In Sec.\ \ref{sec:set} we define the main notions and study the spectrum of boundary modes. The PV renormalization of polarization tensor without boundaries is considered in Sec\ \ref{sec:nb}, where we also discuss the effects due to $v_F\neq 1$. The main part of this work is Sec.\ \ref{sec:PTb} where we formulate the rules of PV scheme with boundaries and compute the renormalized polarization tensor. In Sec.\ \ref{sec:Hall} we compute the Hall conductivity near the boundary and compare to that of a Dirac spinor in $(2+1)$ dimensions. Concluding remarks will be presented in Sec.\ \ref{sec:con}. Some technicalities are contained in Appendices: the parity-odd part of the effcetive action is computed in App. \ref{sec:3D}, while some useful formulas are collected in App.\ \ref{sec:some}.

%%%%%%
\section{The setup}\label{sec:set}
Let us consider one generation of fermions in $(3+1)$ dimensions described by the Dirac operator
\begin{equation}
\slashed{D}=\ii\tilde\gamma^\mu \bigl( \partial_\mu +\ii e A_\mu) +\ii m_5\gamma^5 +m ,\label{Dir}
\end{equation}
where $A_\mu$ is electromagnetic potential. Keeping in mind applications to the condensed matter physics we introduced the Fermi velocity $v_F$ by rescaling the spatial gamma matrices, 
\begin{equation}
\tilde\gamma^\mu=\eta^\mu_\nu \gamma^\nu, \qquad \eta \equiv \mathrm{diag}(1,v_F,v_F,v_F). \label{tilgam}
\end{equation}
We work in the signature $(+---)$, so that before the rescaling $(\gamma^0)^2=1=-(\gamma^a)^2$, $a=1,2,3$. $\gamma^5=-\ii \gamma^0\gamma^1\gamma^2\gamma^3$ is the chirality matrix, so that
\begin{equation}
\mathrm{tr}\, \bigl( \gamma^5\gamma^\mu\gamma^\nu\gamma^\rho\gamma^\sigma \bigr)=4\ii \varepsilon^{\mu\nu\rho\sigma}
\label{trg5}
\end{equation}
with $\varepsilon^{0123}=1$.
 The role of mass parameters $m$ and $m_5$ will be clarified below. 

We assume that the fermions can propagate in a half-space $x^1>0$. Let us introduce two complimentary projectors
\begin{equation}
\Pi_{\pm}=\tfrac 12 (1\mp \ii\gamma^1),\qquad \chi=\Pi_+-\Pi_-=-\ii\gamma^1 \label{Pipm}
\end{equation}
and define the bag boundary conditions \cite{Chodos:1974je,Chodos:1974pn} as
\begin{equation}
\Pi_-\psi(x)\vert_{x^1=0}=0.\label{bcm}
\end{equation}
For the conjugated spinor, $\bar\psi(x)\equiv \psi^\dag(x)\gamma^0$, we have
\begin{equation}
\bar\psi(x)\Pi_+\vert_{x^1=0}=0.\label{bcp}
\end{equation}
These boundary conditions ensure that the normal current vanishes at the boundary, $\bar\psi \gamma^1\psi\vert_{x^1}=0$, and thus provide for the hermiticity of the Dirac hamiltonian. 

Let us describe classical solutions of the free Dirac equation 
\begin{equation}
\slashed{D}_0\psi=0 \label{Dereq}
\end{equation}
 with $\slashed{D}_0\equiv \slashed{D}(A=0)$ subject to bag boundary conditions  \eqref{bcm}. There are oscillating solutions proportional to $e^{\ii k_\mu x^\mu}$ with $k^2=m^2+m_5^2$ that we shall call bulk modes. Other modes, that will be called boundary modes, decay exponentially away from the boundary. To analyze these modes, let us take a particular representation of the gamma matrices in terms of the Pauli matrices $\sigma$:
\begin{equation}
\gamma^0=\begin{pmatrix} 0 & \sigma_1 \\ \sigma_1 & 0 \end{pmatrix},\qquad
\gamma^1= \begin{pmatrix} 0 & \mathds{1} \\ -\mathds{1} & 0 \end{pmatrix},\qquad
\gamma^{2,3}=\ii \begin{pmatrix} 0 & \sigma_{2,3} \\ \sigma_{2,3} & 0 \end{pmatrix}.\label{gams}
\end{equation}
One can easily check that boundary modes have the form
\begin{equation}
\psi_b=e^{mx^1/v_F} \begin{pmatrix} \Psi(x^j) \\ \ii \Psi(x^j) \end{pmatrix}, \quad j=0,2,3,
\label{bmod}
\end{equation}
where the 2-spinors $\Psi$ have to satisfy the Dirac equation in $2+1$ dimensions
\begin{equation}
\bigl( \ii \sigma_1\partial_0 -v_F\sigma_2\partial_2-v_F\sigma_3\partial_3 +m_5\bigr) \Psi=0\,. \label{bDir}
\end{equation}
Thus, boundary modes exist for $m<0$ only, while their mass is given by $m_5$.

We see, that the usual mass $m$ fails to give a gap to all modes since the boundary modes remain gapless. On the contrary, the chiral mass $m_5$ gives a gap to all modes, and this gap tends to infinity for $m_5\to \pm \infty$. This suggests, that it is $m_5$ rather than $m$ that has to be used for the Pauli-Villars subtraction if a boundary is present. A similar observation was made in \cite{Beneventano:2018tle} for graphene nanoribbons, and a similar remedy was suggested.

In this work we are interested in the one-loop effective action for fermions truncated to the 2nd order in external electromagnetic field, 
\begin{equation}
S_{\rm eff}=\frac {\ii e^2}2 \mathrm{Tr}\, \left[ \tilde\gamma^\mu A_\mu \slashed{D}_0^{-1} \tilde\gamma^\nu A_\nu \slashed{D}_0^{-1}\right] \,.\label{Seff2}
\end{equation}
or, in more simple words, in a fermion loop with two photon legs. The Green's function has to satisfy
\begin{equation}
\slashed{D}_{0,x} \ \slashed{D}^{-1}_0(x,y) = \mathds{1} \delta(x-y)\,\qquad \Pi_- \slashed{D}^{-1}_0(x,y)\vert_{x^1=0}=0,\qquad
\slashed{D}^{-1}_0(x,y) \Pi_+\vert_{y^1=0}=0.\label{prop-def}
\end{equation}
To construct this Green's function we notice that 
\begin{equation}
\overline{\sD}_0\sD_0=-\tilde\partial^{\mu}\tilde\partial_\mu - \mfm^2\equiv \Box_\mfm \label{overD} 
\end{equation}
with
\begin{equation}
\overline{\sD}_0=\ii\tilde\gamma^\mu  \partial_\mu  +\ii m_5\gamma^5 -m, \qquad \mfm = \sqrt{m^2+m_5^2} .\label{overD2}
\end{equation}
If there are no boundaries, $\Box_\mfm$ can be easily inverted,
\begin{equation}
G_0(x-y,\mfm)\equiv (\Box_\mfm)^{-1}_{x,y} =  \int \frac{d^4 k}{(2\pi)^4}\, \frac{e^{+\ii k (x-y)}}{ \tilde k^2 -\mfm^2 + \ii 0 }
=-\frac{\ii \mfm}{4\pi^2 v_F^3}\frac{\mathrm{K}_1(\mfm \sqrt{{-\lambda + \ii 0}})}{\sqrt{{-\lambda + \ii 0}}}, \label{scprop}
\end{equation}
where
\begin{equation}
\lambda  =  (x^0 - y^0)^2 - v_F^{-2}(x^a - y^a)^2,\qquad a=1,2,3,
\end{equation}
and $\mathrm{K}_1(z)$ is the modified Bessel function.

Let $x^\|$ be a projection of vector $x$ to the boundary plane, and let $\bar x$ denote a reflected vector, $x^\|=\bar x^\|$ and $x^1=-\bar x^1$. Then the full propagator in coordinate representation reads
\begin{equation}
\slashed{D}_0^{-1}(x,y)=\overline{\sD}_{0,x} \bigl( G_0(x-y,\mfm ) - \chi G_0(x-\bar y,\mfm )
+2\Pi_- \Gh (x-\bar y,m,m_5) \bigr) \,, \label{coorprop}
\end{equation}
where
\begin{equation}
\Gh (x-\bar y,m,m_5)=-\frac m{v_F}\int_0^\infty dz e^{-zm/v_F} G_0(x+z_*-\bar y,\mfm). \label{defGh}
\end{equation}
Here $z_*$ is a vector such that $z_*^1=z$ and $z_*^\|=0$. Equations (\ref{prop-def}) are checked by inspection. In what follows, we shall drop the masses from the notations whenever this cannot lead to a confusion.

There is an important observation regarding the dependence of propagator in coordinate representation on the Fermi velocity. To obtain (\ref{coorprop}), it is sufficient to take the full propagator with $v_F=1$ and make the replacement
\begin{equation}
x^0,\,y^0 \to v_Fx^0,\, v_Fy^0,\quad m\to m/v_F,\quad m_5\to m_5/v_F.\label{vF1}
\end{equation}
This may be verified directly or demonstrated on general grounds.

For the future use we define a Minkowski norm for space-time vectors as
\begin{equation}
|x|:=\sqrt{ -g_{\mu\nu}x^\mu x^\nu +\ii 0}\,,\label{Mink}
\end{equation}
where $g=\mathrm{diag}(+1,-1,-1,-1)$. The sign under square root is chosen to simplify the Wick rotation, while $\ii 0$ governs the phase. For Wick rotated vectors $x^0\to -\ii x^4$, and the norm is defined in the usual way, 
$|x_E|=\sqrt{(x^1)^2+(x^2)^2+(x^3)^2+(x^4)^2}$.

\section{No-boundary case}\label{sec:nb}
In this Section we compute the polarization tensor in Minkowski space without boundaries. The computations are rather standard, though there are two important differences: the presence of $v_F$ and of both axial and normal masses. We start with
\begin{equation}
\Pi^{\mu\nu}(p) = \frac{\ii e^2}{(2\pi)^4} \int   {d^4k}  \,
\mathrm{tr}
  \left[ 
    \tga^\mu \sD_0^{-1}(k) \tga^\nu \sD_0^{-1}(k -p)
  \right], \label{Pmn42}
\end{equation}
where
\begin{equation}
\sD_0^{-1}(k)  =   \frac{ - k_\mu \tga^\mu + i m_5 \gamma^5 -m  }{ \tilde k^2-m_ 5^2 -m^2}.
\end{equation}
After taking the trace and making the change of the integration variable $k_\mu\to \tilde k_\mu =\eta_\mu^\nu k_\nu$ in (\ref{Pmn42}) we make an important observation:
\begin{equation}
\Pi^{\mu\nu}(p) =v_F^{-3} \
		\eta^\mu_\alpha \widehat\Pi^{\alpha\beta}(\tilde p) \eta^\nu_\beta \,, \label{vFinPi}
\end{equation}
where $\widehat \Pi$ is the usual polarization tensor computed for $v_F=1$ with the ordinary mass equal to $\mfm$ and no chiral mass parameter. The computation of $\widehat \Pi$ goes as in textbooks, see e.g. \cite{Bogoliubov}. After performing the Wick rotation and introducing the Feynman parameters, we arrive at the integral
\begin{equation}
\widehat \Pi^{\mu\nu}(k) =-\frac{e^2}{2\pi^2} (k^\mu k^\nu -g^{\mu\nu}k^2)\int_0^1dx\int_0^\infty d\alpha \frac{x(x-1)}{\alpha} e^{-\alpha(x(1-x)k^2+\mfm^2)}. \label{Pmn51}
\end{equation}
The integral over $\alpha$ is divergent at the lower limit. To make the $\widehat\Pi$ finite, it is sufficient to add two PV regulators with weights $c_i$ and $\mfm^2$ replaced by $M_{i}^2=m_i^2+m_{5,i}^2$, $i=1,2$ satisfying the conditions\footnote{In the no-boundary case the polarization tensor depends on the masses only in the combinations $\mfm^2$ (or $M_{i}^2$). Thus, there is no need to introduce axial masses here. These masses, however, will be essential in the presence of a boundary.}
\begin{equation}
1+c_1+c_2=0,\qquad \mfm^2+c_1M_1^2+c_2M_2^2=0 \,.\label{PVcond}
\end{equation}
Finally, after returning to the Minkowski signature, we obtain
\begin{equation}
\widehat \Pi^{\mu\nu}(p) =
- \frac{ e^2}{2\pi^2} \bigl[p^\mu p^\nu -g^{\mu\nu}p^2\bigr]
\left( \frac {c_1 \ln (M_1^2	/ \mfm^2)}6+\frac {c_2 \ln (M_2^2/ \mfm^2 )}6 + 
	\int_0^1dx\, x(1-x) \ln \left[ 1-\frac{x (1-x)p^2}{\mfm^2}\right]
	\right)\,,
\label{Pi ord ren}
\end{equation}
where we dropped the terms that vanish in the limit $M_{1,2}^2\to\infty$. The terms with $\ln M_i^2$ are divergent at this limit, and these divergences have to be removed by suitable counterterms. 

Counterterms needed to renormalize a theory should all be local expressions having correct invariance properties and correct canonical mass dimensions. Since all quasi-relativistic symmetries are broken by the presence of different characteristic velocities for fermions and photons, the allowed counterterms depending just on the electromagnetic field have the form of the Maxwell action in a media\footnote{Since the parity invariance has been already violated by the presence of $\gamma^5$ in the Dirac operator, a term $\vec E \cdot \vec B$ is also allowed. This term is a total derivative and thus is not essential on $\mathbb{R}^4$. In the presence of a boundary, however, this term leads to a shift of the Chern-Simons coupling. The resulting ambiguity is removed by requiring that the effective action vanishes in the limit of an infinite mass gap, $|m_5|\to\infty$. In this way, one recovers the results reported in Sections \ref{sec:nb} and \ref{sec:Hall}. We shall not return to this issue any more.}
\begin{equation}
S_{\rm EM}=\frac 12 \int d^4x \left( \epsilon \vec{E}^2 - \frac 1\mu \vec{B}^2 \right) =
\frac 12 \int \frac{d^4p}{(2\pi)^4} \left( \epsilon \vec{E}(-p)\cdot \vec E(p) -
\frac 1\mu \vec B(-p)\cdot \vec{B}(p)\right)\,. \label{EMclass}
\end{equation}

Let us write the one-loop effective action following from (\ref{Pi ord ren}) as
 {\begin{eqnarray}
&&S_{\rm eff}=\frac 12 \int \frac{d^4p}{(2\pi)^4} A_\mu(-p) \Pi^{\mu\nu}(p) A_\nu(p) \nonumber\\
&&\quad = -\frac 12 \int \frac{d^4p}{(2\pi)^4} \, \frac{e^2}{2\pi^2} \left( v_F^{-1} \vec{E}(-p)\cdot \vec E(p) -
v_F \vec B(-p)\cdot \vec{B}(p)\right) \left( \frac{c_1}{6} \ln \frac {M_1^2}{\mfm^2} 
+ \frac{c_2}{6} \ln \frac {M_2^2}{\mfm^2} +f(\tilde p^2/\mfm^2) \right), \label{S11}
\end{eqnarray} }
Here we restored the $v_F$ dependence according to (\ref{vFinPi}), and defined
 {\begin{equation}
f(z)\equiv \int_0^1dx\, x(1-x) \ln (1-{x (1-x)}z)= 
	 \frac{-\sqrt{z} (12+5 z)+ 6 (z+2)\sqrt{4-z}\mathop{\rm arctan}\sqrt{z/(4-z)}}{18 z^{3/2}}, \quad z\in(0,4)\,.
	 \label{fofz}
\end{equation} }
An analytical continuation to other values of $z$ is assumed when necessary. 

The divergences in (\ref{S11}) are canceled  by the following renormalization of $\epsilon$ and $\mu$ 
 {\begin{eqnarray}
&&\delta_1 \epsilon =\frac {e^2}{2\pi^2 v_F} \left( \frac{c_1}{6} \ln \frac {M_1^2}{\mfm^2} 
+ \frac{c_2}{6} \ln \frac {M_2^2}{\mfm^2} \right) + \mbox{finite} \label{del1eps}\\
&&\delta_1 \frac 1\mu =\frac {e^2 v_F}{2\pi^2} \left( \frac{c_1}{6} \ln \frac {M_1^2}{\mfm^2} 
+ \frac{c_2}{6} \ln \frac {M_2^2}{\mfm^2} \right) + \mbox{finite} . \label{del1mu}
\end{eqnarray}}
Finite parts in (\ref{del1eps}) and (\ref{del1mu}) have to be fixed by a suitable normalization condition. We request that the kernel of (\ref{S11}) jointly with contributions from the counterterms vanishes when $\tilde p^2=\lambda^2$ for some scale $\lambda$. The renormalized one-loop effective action becomes
 {\begin{equation}
S_{\rm eff}^{\rm ren}= -\frac 12 \int \frac{d^4p}{(2\pi)^4} \, \frac{e^2}{2\pi^2} \left( v_F^{-1} \vec{E}(-p)\cdot \vec E(p) -
v_F \vec B(-p)\cdot \vec{B}(p)\right) (f(\tilde p^2/\mfm^2) - f(\lambda^2/\mfm^2)) .\label{S1ren}
\end{equation}} 
This means that $\epsilon$ and $\mu$ in the classical action (\ref{EMclass}) have the values measured for the photons with $\tilde p^2=\lambda^2$. Since physics cannot depend on the choice of $\lambda$, the scale dependence of the dielectric constant and magnetic permeability is defined by $f(\lambda^2/\mfm^2)$. However, in a full theory of quantized photons and fermions all parameters ($e$, $v_F$, etc.) become scale dependent, as dictated by the Renormalization Group equations, see \cite{Gonzalez:1993uz}. A single computation of the polarization tensor is not enough to fix the running of $\epsilon$ and $\mu$, but one can draw some qualitative conclusions regarding this running already here. First of all, due to the presence of $v_F^2$ in $\tilde p^2$, the dependence of $\epsilon$ and $\mu$ on the spatial momenta is very small. The amplitude of quantum corrections to $\epsilon$ is of the order $e^2/v_F$, while to $\mu$ -- of the order of $e^2v_F$. Therefore, we expect that the scale dependence of $\epsilon$ to be of the order of unity, while the scale dependence of $\mu$ to be negligible. Qualitatively, all these conclusions are consistent with what we know about dielectric properties of the bulk of Topological Insulators.

One can easily check that $S_{\rm eff}^{\rm ren}$ vanishes in the limit $\mfm^2\to\infty$ and is regular at $\tilde p^2\to 0$ and at $\mfm^2\to 0$. 

%%%%%%
\section{Polarization tensor in the presence of a boundary}\label{sec:PTb}
\subsection{Unregularized expressions}\label{sec:unreg}
In the presence of a boundary, it is convenient to work in the coordinate representation. The effective action (\ref{Seff2}) reads
\begin{eqnarray}
&&S_{\rm eff}=  \frac{\ii e^2}{2} \,  \int d^4x\int  d^4y\, A_{\mu}(x)A_{\nu}(y)\, 
\mathrm{tr}\left(\tilde\gamma^{\mu}\slashed{D}_0^{-1}(x,y)\tilde\gamma^{\nu}\slashed{D}_0^{-1}(y,x)\right)\nonumber\\
&&\qquad \equiv \frac 12 \,  \int d^4x\int  d^4y\, A_{\mu}(x)A_{\nu}(y)\, \Pi^{\mu\nu}(x,y)
,  \label{SeffCoord}
\end{eqnarray}
where the integration runs over the half-space $\mathbb{R}_+\times \mathbb{R}^3$. The propagator $\slashed{D}_0^{-1}$ has been defined in eq.\ (\ref{coorprop}).

Again, there are simple rules to reintroduce $v_F$ in the polarization tensor. One has to take the tensor $\widehat \Pi$ computed with $v_F=1$, contract it with $\eta$ (\ref{tilgam}), and make the replacement as in Eq.\ (\ref{vF1}). Symbolically,
\begin{equation}
\Pi^{\mu\nu}(x,y)=\eta_\alpha^\mu \eta_\beta^\nu \widehat\Pi^{\alpha\beta}(x,y)\vert_{\mathrm{Eq.\ (\ref{vF1})}}. \label{vF12}
\end{equation}
Note, that this rule differs from (\ref{vFinPi}) that we used in the Fourier representation.

To compute the trace in (\ref{SeffCoord}) it is convenient to split the propagator as
\begin{eqnarray}
&&\slashed{D}_0^{-1}(x,y) = \overline{\sD}_{0,x} \bigl( G_1(x,y)-\chi G_2(x,y)\bigr),\nonumber\\
&&G_1(x,y)=G_0(x-y)+\Gh(x-\bar y),\qquad G_2(x,y)=G_0(x-\bar y)+\Gh(x-\bar y). \label{GGG}
\end{eqnarray}
Now the terms under the trace in (\ref{SeffCoord}) can be separated in two groups: the ones containing an even number of gamma matrices and the ones containing an odd number of them (recall that $\chi=-\ii\gamma^1$). According to this separation, we represent
\begin{equation}
S_{\rm eff}=S_{\rm even}+S_{\rm odd} .\label{evenodd}
\end{equation} 
We shall call these parts parity even ($P$-even) and parity odd ($P$-odd), respectively. The parity transformation is understood as an inversion of orientation of the space-time resulting in an inversion of the sign in front of the Levi-Civita tensor in (\ref{trg5}).

The polarization tensor of $P$-even part reads
\begin{eqnarray}
&&\Pi_{\mathrm{even}}^{\mu\nu}(x,y) = 4\ii e^2\big(-  T^{\mu\lambda\nu\xi}\,\partial_{\lambda[x]}G_1(x,y)\cdot\partial_{\xi[y]}G_1(y,x) + g^{\mu\nu}\left(m_5^2 + m^2\right)\left(G_1(x,y)\right)^2 \nonumber\\
&&\qquad\qquad - \overline{T}^{\mu\lambda\nu\xi}\,\partial_{\lambda[x]}G_2(x,y)\cdot\partial_{\xi[y]}G_2(y,x) + \bar{g}^{\mu\nu}\left(m_5^2 - m^2\right)\left(G_2(x,y)\right)^2 \nonumber\\
&&\qquad\qquad + m\left[T^{\mu\nu\xi 1} G_1(x,y)\cdot\partial_{\xi[y]}G_2(y,x) + T^{\mu 1 \nu\xi } G_2(x,y)\cdot\partial_{\xi[y]}G_1(y,x)+(\mu\leftrightarrow\nu, \, x\leftrightarrow y)\right]  \big),
\label{evenpolop}
\end{eqnarray}
where $\bar g$ is the Minkowski metric with a reflected $(1,1)$ component,
\begin{equation}
\bar g=\mathrm{diag}\,\left(+1,+1,-1,-1\right) \label{tildmetr}
\end{equation}
and
\begin{eqnarray}
&&T^{\mu\lambda\nu\xi}  = 
 g ^{\lambda\mu} g ^{\xi\nu} -  g ^{\lambda\xi} g ^{\mu\nu}  +  g ^{\lambda\nu} g ^{\xi\mu}, \nonumber\\
&&\overline{T}^{\mu\lambda\nu\xi}  = 
{ g }^{\lambda\mu}{ g }^{\xi\nu} - \bar{ g }^{\lambda\xi}\bar{ g }^{\mu\nu}  + \bar{ g }^{\lambda\nu}\bar{ g }^{\xi\mu} 
. \label{vartraces}
\end{eqnarray}

The parity odd part is
\begin{equation}
\Pi_{\mathrm{odd}}^{\mu\nu}(x,y) =  -\ii \varepsilon^{1\mu \rho \nu} \,\partial_{\rho [y]} \mathrm{Q}_{\mathrm{4}}(x,y), \label{oddpolop}
\end{equation}
It corresponds to the effective action
\begin{equation}
S_{\mathrm{odd}}
= \frac{\ii}{2}\int d^4xd^4y \left(\varepsilon^{1i j k}\, A_{i}(x) \,\partial_{j[y]}  A_{k}(y)\right) \cdot \mathrm{Q}_{4}(x,y) , \label{2mSodd}
\end{equation} 
with the form factor 
\begin{eqnarray}
&& \mathrm{Q}_4(x,y)=8m_5e^2 G_1(x,y)G_2(x,y)\nonumber\\
&&\qquad\qquad =8m_5e^2 \left( G_0(x-y)G_0(x-\bar y) +G_0(x-y)\Gh (x-\bar y) + G_0(x-\bar y)\Gh(x-\bar y) +
\Gh(x-\bar y)^2\right) .\label{G12expl}
\end{eqnarray}

%%%%%%%
\subsection{Pauli-Villars regularization and finiteness}\label{sec:PV}
Below we analyze the ultraviolet (short distance) singularities of the polarization tensor in the Euclidean region. As in the boundaryless case we work with two PV regulators, defining the regularized polarization operator as follows: 
\begin{equation}
	[\Pi^{\mu\nu}]_{\rm reg} := \Pi^{\mu\nu}(m,m_5) + \sum_{i=1}^2 c_i \Pi^{\mu\nu}(m_i,m_{5,i}).
	%\quad  M_{i}^2=m_i^2+m_{5,i}^2
	\label{[]reg}
\end{equation}We notice, however, that  it depends separately on $m$ and $m_5$, unlike  \eqref{Pmn51}.

The singularities of various constituents of $P$-even  \eqref{evenpolop} and and $P$-odd  \eqref{oddpolop} polarization tensors are described by Eqs.\ (\ref{G0SingStruct}) and (\ref{Gestim}).
It can be shown now that in $[\Pi^{\mu\nu}_{\mathrm{even}}]_{\rm reg}$ almost all non-integrable singularities disappear under the conditions (\ref{PVcond}). However, there still remain the non-integrable singularities of the types 
\begin{equation}
u_+^{-2}u_-^{-3},\quad u_+^{-3}u_-^{-2},\quad {\rm and}\quad u_+^{-5} \label{types}
\end{equation}
with 
\begin{equation}
u_-=|x_E-y_E|,\qquad u_+=|x_E-\bar y_E|\,.\label{ump}
\end{equation}
The coefficients in front of these singularities are proportional to the ordinary mass, so that their cancellation requires an additional condition
\begin{equation}
m+c_1m_1+c_2m_2=0.\label{req3}
\end{equation}

In its turn, the parity odd polarization tensor (\ref{oddpolop}) contains a dangerous singularity proportional to $u_-^{-3}u_+^{-2}$. However, in the effective action this tensor is multiplied by an antisymmetric combination $\epsilon^{1\mu\rho\nu} A_\mu (x)A_\nu(y)$ that vanishes in the coincidence limit. Thus this singularity becomes milder and does not lead to any divergence.

We see, that in the presence of a boundary the conditions (\ref{PVcond}) have to be supplemented by an additional condition (\ref{req3}). These three equations admit solutions with arbitrarily large axial masses of the PV fields. For example, one can take $m=m_1=m_2$ and axial masses satisfying $m_5^2 + c_1m_{5,1}^2+c_2m_{5,2}^2=0$, which resembles the second condition in (\ref{PVcond}). There are, of course, other solutions as well, but we shall not rely on any particular choice.

Summarizing, our PV prescription is as follows. We take two PV regulators with the weights $c_1$, $c_2$, masses $m_1$ and $m_2$, and axial masses $m_{5,1}$ and $m_{5,2}$. We impose the restrictions  \eqref{PVcond} and  \eqref{req3} on the weights and masses. 
The physical limit corresponds to infinite axial masses of the PV regulators, whilst the ordinary masses, which, we remind, do not give mass gaps to the surface modes, are kept finite. As we shall see, the renormalized effective action will not depend on a particular choice  of a solution of (\ref{PVcond}) and (\ref{req3}). We also take the axial masses $m_{5,i}$ of the same sign as $m_5$ for the reason that will become clear in Sec. \ref{sec:renO}.

Let us make an important remark.  The conditions (\ref{PVcond}) and (\ref{req3}) do not admit a solution with $|m_{1}|,\, |m_2|\to \infty$ and finite axial masses of PV regulators. On the other side, such a limit would be the only reasonable opportunity if we did not introduce the axial masses in this game. Thus, \emph{the usual PV scheme, which does not rely on the axial masses, fails to give a finite result in the presence of a boundary.}

In $2+1$ dimensions the situation is different. In the case of a single generation of Dirac fermions there is no second mass term. However, since the divergences are milder, the polarization tensor can be computed (presumably) even without any explicit regularization \cite{Fosco:2018nio}. (The paper \cite{Fosco:2018nio} gives too few details to make more definite statements). The problems may arise if one attempts to move the results of \cite{Fosco:2018nio} closer to condensed matter applications by making a PV subtraction with respect to the usual mass. As has been demonstrated in \cite{Beneventano:2018tle} for nanoribbons, the contribution of edge states to the conductivity is treated incorrectly in such a procedure. The same problem may persist in half-space as well.

In conclusion we notice that the regularized effective action is gauge invariant, as expected. This provides a useful cross-check for our approach, which is, however, too long and too technical to be reported here.  In what follows we address its renormalization i.e. consider the physical limit $|m_{5,i}|\to\infty$.
%%%%%%

\subsection{Renormalization of the parity even part}\label{sec:renE}

It is convenient to split ${\Pi_{\mathrm{even}}^{\mu\nu}}$ in three parts:
\begin{equation}
\Pi_{\mathrm{even}}^{\mu\nu}(x,y) = \Pi^{\mu\nu}_\mathrm{bulk}(x,y)  + \Pi^{\mu\nu}_\mathrm{mirr}(x,y) +\Pi^{\mu\nu}_\mathrm{rest}(x,y),
\end{equation}
where
\begin{eqnarray}
&&\Pi^{\mu\nu}_\mathrm{bulk}(x,y)=4\ii e^2\Big(-  T^{\mu\lambda\nu\xi}\,\partial_{\lambda[x]}G_0(x,y)\cdot\partial_{\xi[y]}G_0(y,x) + g^{\mu\nu}\mathfrak{m}^2\left(G_0(x,y)\right)^2 \Big),\nonumber\\
&&\Pi^{\mu\nu}_\mathrm{mirr}(x,y)=4\ii e^2\Big(- \overline{T}^{\mu\lambda\nu\xi}\,\partial_{\lambda[x]}G_0(x,\bar{y},\mathfrak{m})\cdot\partial_{\xi[y]}G_0(x,\bar{y}) + \bar{g}^{\mu\nu}\mathfrak{m}^2\left(G_0(x,\bar{y})\right)^2 \Big).
\label{Pbm} 
\end{eqnarray}
The tensor $\Pi_{\rm bulk}$ is obtained from the first line on the right hand side of (\ref{evenpolop}) by keeping only $G_0(x,y)$ in $G_1(x,y)$. To obtain $\Pi_{\rm mirr}$ one has to keep the terms in $G_2(x,y)$ containing $G_0(x,\bar y)$ together with one extra term, that converts $-m^2$ to $m^2$. Both $\Pi_{\rm bulk}$ and $\Pi_{\rm mirr}$ can be represented through a single form factor
\begin{eqnarray}
&&\Pi^{\mu\nu}_\mathrm{bulk}(x,y)=\frac{4\ii e^2}{(2\pi)^4}\left(\partial^{\mu[x]}\partial^{\nu[x]} - \partial_{[x]}^2 g ^{\mu\nu}\right)\, \mathfrak{m}^4\,\mathcal{P}(u_{-}\mathfrak{m} ) , 
\nonumber\\
&&\Pi^{\mu\nu}_\mathrm{mirr}(x,y)=\frac{4\ii e^2}{(2\pi)^4} \left(\partial^{\mu[x]}\bar{\partial}^{\nu[x]} - \partial_{[x]}^2\bar{ g }^{\mu\nu}\right) \,\mathfrak{m}^4\,\mathcal{P}(u_{+} \mathfrak{m}) \label{ginvPolOp}
\end{eqnarray}
with 
\begin{equation}
\mathcal{P}(z) := \frac{   \left( {z}^{2}-1 \right)  \left( 
{{\rm K}_1\left(z\right)} \right) ^{2} - z \,
{{\rm K}_1\left(z\right)}{{\rm K}_0\left(z\right)} -   {z}^{2}\left( 
{{\rm K}_0\left(z\right)} \right) ^{2} }{3{{z}^{2}}}. \label{Pdef}
\end{equation}

Let us consider the effective action corresponding to $\Pi_{\mathrm{bulk}}$ and $\Pi_{\mathrm{mirr}}$. After integration by parts it can be written in the form
\begin{equation}
\frac{1}{2} \int d^4 x d^4y\, A_{\mu}(x) \,\left[\Pi^{\mu\nu}_{\mathrm{bulk}}(x,y) +\Pi^{\mu\nu}_{\mathrm{mirr}}(x,y)\right]_{\mathrm{reg}}\,  A_{\nu}(y) =
 \left(S_{\mathrm{bulk}}[A,\mathfrak{m}]\right)_{\mathrm{reg}} + \left(S_{\mathrm{mirr}}[A,\mathfrak{m}]\right)_{\mathrm{reg}},
\end{equation}
where
 {\begin{eqnarray}
&&\left(S_{\mathrm{bulk}}[A]\right)_{\mathrm{reg}} = {\frac{\ii e^2 }{(2\pi)^4}} \int d^4xd^4y\,
%\left(
F_{\mu\nu}(x) F^{\mu\nu}(y) \left[\mathfrak{m}^4\mathcal{P}(u_{-}\mathfrak{m} )\right]_{\mathrm{reg}} %\right)
, \nonumber\\
&&\left(S_{\mathrm{mirr}}[A]\right)_{\mathrm{reg}} = {\frac{\ii e^2 }{(2\pi)^4}} \int d^4xd^4y\, 
%\left(
\overline{F}_{\mu\nu}(x) {F}^{\mu\nu}(y) \left[\mathfrak{m}^4\mathcal{P}(u_{+}\mathfrak{m} )\right]_{\mathrm{reg}}
%\right)
.
\label{actginvreg}
\end{eqnarray} }
Here we defined
\begin{equation}
\overline{F}_{\mu\nu}(x) = \bar{\partial}_{\mu}\bar{A}_{\nu}(x) -  \bar{\partial}_{\nu}\bar{A}_{\mu}(x),\quad\quad
 \bar{A}_{\nu}(x)\equiv \bar{g}^{\xi}_{\nu}{A}_{\xi}(x).
\end{equation}
Note, that the surface terms produced by integration by parts are canceled in $S_{\mathrm{bulk}}+S_{\mathrm{mirr}}$ but not in each of them separately. Both $S_{\rm bulk}$ and $S_{\rm mirr}$ are manifestly gauge invariant.

The  action $S_{\mathrm{bulk}}$  depends on the boundaries through the integration region only. It can be obtained from (\ref{S11}) by computing the Fourier integral of the kernel and then restricting the ranges of coordinates to $x^1,y^1\geq 0$. The renormalization thus goes exactly the same way as has been explained in Sec.\ \ref{sec:nb} (we checked), though the computations are much more complicated in the coordinate representation. 
We do not present any details here.

Let us turn to $S_{\mathrm{mirr}}$. This contribution to the effective action describes interaction of the electromagnetic field with a ``mirror" current. Note, that
\begin{equation}
\mathcal{P}(z) \simeq  -\frac{1}{3 z^4} + \mathcal{O}\left(z^{-2}\right), \quad \mbox{at} \quad z \longrightarrow 0.
\end{equation}
The singularity $u_+^{-4}$ is integrable in the half-space. Thus $S_{\mathrm{mirr}}$ does not require any regularization by itself. However, the PV subtraction may be non-trivial. To study the $|m_5|\to\infty$ limit, let us change the integration variables as
\begin{equation}
x^\|-y^\|=v^\| |\mfm|^{-1},\quad y^\|=w^\|,\quad x^1=v^1 |\mfm|^{-1},\quad y^1=w^1|\mfm|^{-1}.\label{chvar1}
\end{equation}
Then,
 {\begin{equation}
\mfm \cdot S_{\rm mirr}=\frac {\ii e^2}{(2\pi)^4} \int d^4 v\, d^4w \overline{F}_{\mu\nu}\left( w^\| +\frac{v^\|}{\mfm},\frac {v^1}{\mfm} \right)\, F_{\mu\nu}\left( w^\|,\frac{w^1}{\mfm}\right)\, \mathcal{P}(|v^\|,v^1+w^1|).\label{0144}
\end{equation} }
Obviously, the limit $\mfm\to\infty$ of the right hand side of (\ref{0144}) is finite. Consequently,
\begin{equation}
\lim_{\mfm\to \infty} S_{\rm mirr}=\lim_{|m_5|\to \infty} S_{\rm mirr}=0.\label{0147}
\end{equation}
Thus, the PV subtraction does not change the expression (\ref{actginvreg}) for $S_{\rm mirr}$ and
\begin{equation}
S_{\rm mirr}^{\rm ren}=S_{\rm mirr}.\label{Smirrren}
\end{equation}

It remains to renormalize $\Pi_{\rm rest}$. In the regularized expression 
\begin{equation}
\left[ \Pi_{\rm rest}^{\mu\nu}(x,y,m,m_5)\right]_{\rm reg} \equiv \Pi_{\rm rest}^{\mu\nu}(x,y,m,m_5) +\sum_{i=1}^2 c_i
\Pi_{\rm rest}^{\mu\nu}(x,y,m_i,m_{5,i}) \label{restreg}
\end{equation}
all singularities are integrable if (\ref{PVcond}) and (\ref{req3}) are satisfied, though each of the individual terms has singularities of the types \eqref{types}. Before taking the limit $|m_{5,i}|\to\infty$, let us isolate these singularities (which will allow us to treat the terms in (\ref{restreg}) separately). To this end, we rewrite the corresponding effective action as
\begin{equation}
\left[S_{\mathrm{rest}}\right]_{\mathrm{reg}} =  \int d^3 z^{\|}\int_0^{\infty}d x^1 \int_0^{\infty} d y^1 \,\Phi_{\mu\nu}(z^{\|},x^{1},y^{1})  \,\left[\Pi^{\mu\nu}_{\mathrm{rest}}(z^{\|}, x^1,y^1,m,m_5)\right]_{\mathrm{reg}},  \label{hardeffactQua}
\end{equation}
where we introduced a new integration variable $z^\| = x^{\|} - y^{\|}$ and defined
\begin{equation}
\Phi_{\mu\nu}(z^{\|},x^{1},y^{1}) = \frac{1}{2} \int d^3 y^{\|} A_{\mu}(y^{\|} + z^{\|},x^1)A_{\nu}(y^{\|},y^1).
\end{equation}
Now, we add and subtract the term $\Phi(0,0,0)$ under the integral in (\ref{hardeffactQua}).
\begin{eqnarray}
&&\left[S_{\mathrm{rest}}\right]_{\mathrm{reg}} =  \int d^3 z^{\|}\int_0^{\infty}d x^1 \int_0^{\infty} d y^1 \,\left(\Phi_{\mu\nu}(z^{\|},x^{1},y^{1}) -\Phi_{\mu\nu}(0,0,0) \right) \,\left[\Pi^{\mu\nu}_{\mathrm{rest}}(z^{\|}, x^1,y^1,m,m_5)\right]_{\mathrm{reg}}\nonumber\\
&&\qquad\qquad +\int d^3 z^{\|}\int_0^{\infty}d x^1 \int_0^{\infty} d y^1 \,\Phi_{\mu\nu}(0,0,0)  \,\left[\Pi^{\mu\nu}_{\mathrm{rest}}(z^{\|}, x^1,y^1,m,m_5)\right]_{\mathrm{reg}}\label{0210}
\end{eqnarray}
The second line in (\ref{0210}) vanishes. This follows from the following facts. First, one may represent 
\begin{equation}
\Phi_{\mu\nu} (0,0,0) = \partial_{\mu[w]} \left(w^{\lambda}\cdot\Phi_{\lambda\nu} (0,0,0)\right), \quad\quad w^j \equiv z^j,\quad w^1 \equiv x^1.
\end{equation}
Second, the regularized $\Pi_{\rm rest}$  e{does not contain problematic singularities, what}  allows us to integrate by parts. Third, direct calculations show that $[\Pi_{\rm rest}]_{\rm reg}$ is transversal and satisfies the conditions $[\Pi_{\rm rest}^{1\mu}]_{\rm reg}|_{x^1 =0} = 
0 = [\Pi_{\rm rest}^{\mu1}]_{\rm reg}|_{y^1=0}$, that guarantee the absence of boundary terms upon integration by parts. 

The combination $\Phi_{\mu\nu}(z^{\|},x^{1},y^{1}) -\Phi_{\mu\nu}(0,0,0)$ vanishes at the point $u_-=u_+=0$ where the unregularized polarization tensor has a nonintegrable singularity. The singularity of integrand on the first line of (\ref{0210}) becomes milder, so that the contributions of the physical field and of each of the regulators become finite. Let us consider a contribution of one of the regulator fields. After a rescaling of the coordinates with $|m_{5,i}|$, we obtain
\begin{equation}
|m_{5,i}|\int d^3 z^{\|}\int_0^{\infty}d x^1 \int_0^{\infty} d y^1 \,\left(\Phi_{\mu\nu}\left(\frac{z^{\|}}{|m_{5,i}|},\frac{x^{1}}{|m_{5,i}|},\frac{y^{1}}{|m_{5,i}|}\right) -\Phi_{\mu\nu}(0,0,0) \right) \,\Pi^{\mu\nu}_{\mathrm{rest}}(z^{\|}, x^1,y^1,m_i/|m_{5,i}|,1). \label{1610}
\end{equation}
The difference of two $\Phi_{\mu\nu}$ terms behaves as $|m_{5,i}|^{-1}$ at large $|m_{5,i}|$. All terms in the polarization tensor $\Pi_{\rm rest}$ contain either a factor of $m$ or at least one $\Gh$ which is proportional to the mass. Thus the rescaled $\Pi_{\rm rest}$ behaves as $m_i/|m_{5,i}|$. Therefore, we conclude that (\ref{1610}) vanishes in the limit $|m_{5,i}|\to\infty$, and then
\begin{equation}
S_{\mathrm{rest}}^{\mathrm{ren}}=\lim_{|m_{5,i}|\to \infty}\left[S_{\mathrm{rest}}\right]_{\mathrm{reg}}
=\frac 12 \int d^4x\, d^4y \, \left( A_\mu(x)A_\nu(y) - A_\mu(x^\|,0)A_\nu(x^\|,0) \right) \Pi_{\rm rest}^{\mu\nu}(x,y,m,m_5) .\label{Srestren}
\end{equation} 
It is easy to see, that $\lim_{m_5\to \infty} S_{\mathrm{rest}}^{\mathrm{ren}}=0$.

We conclude this subsection with a short guide to renormalized expressions for $S_{\rm even}$. It is represented by a sum of three contributions, $S_{\mathrm{bulk}}^\mathrm{ren}+S_{\mathrm{mirr}}^\mathrm{ren}+S_{\mathrm{rest}}^\mathrm{ren}$. The renormalization of bulk part has been performed in Sec. \ref{sec:nb}. The mirror part is given by Eq.\ (\ref{actginvreg}). It does not need any PV subtractions. The last term is (\ref{Srestren}), where the polarization tensor $\Pi_{\rm rest}$ is a rather long expressions defined as a difference between $\Pi_{\rm even}$, Eq.\ (\ref{evenpolop}) and other two tensors, $\Pi_{\rm bulk}$ and $\Pi_{\rm mirr}$, that are presented in (\ref{Pbm}).
%%%
\subsection{Renormalization of the parity odd part.}\label{sec:renO} 
It remains to make the Pauli-Villars subtraction in the parity odd effective action (\ref{2mSodd}). First of all, we expand the notations $\mathrm{Q}_4(x,y)\to \mathrm{Q}_4(x^\|-y^\|,x^1,y^1,m,m_5)$. After changing the variables similarly to (\ref{chvar1}), 
\begin{equation}
x^\|-y^\|=v^\| |m_5|^{-1},\quad y^\|=w^\|,\quad x^1=v^1 |m_5|^{-1},\quad y^1=w^1|m_5|^{-1},
\label{chvar}
\end{equation}
we arrive at the expression
\begin{equation}
S_{\mathrm{odd}} =  \frac{\ii}{2}\,\mathrm{sgn}(m_5) \, \int d^4v\, d^4w \,
\mathrm{Q}_{\mathrm{4}}(v^{\|},v^n,w^n, m/|m_5|,1)\,
 A_{i}\left(w^{\|} + \frac{v^{\|}}{|m_5|}, \frac{v^1 }{|m_5|} \right) \frac{\partial}{\partial w^j}A_{k}\left(w^{\|} , \frac{ w^1 }{|m_5|} \right)
\varepsilon^{1 i j k}. \label{2mstep2}
\end{equation}
In the large $|m_5|$ limit, the contribution from $\mathrm{Q}_4$ can be factored out
\begin{equation}
\lim_{|m_5|\to \infty} S_{\mathrm{odd}}={-} \mathrm{sgn}(m_5)\cdot{\mathcal{C}}\cdot \int d^3 w^{\|}\,
A_{i}\big(w^{\|},0\big) \frac{\partial}{\partial w^j} A_{\nu}\big(w^{\|},0 \big)
\varepsilon^{ni j k} ,\label{factorized}
\end{equation}
where 
\begin{equation}
\mathcal{C}=-\frac{\ii}{2} \int d^3 v^{\|} \int_0^\infty d v^1\int_0^{\infty} dw^1\,\, 
\mathrm{Q}_{\mathrm{4}}(v^{\|},v^1,w^1, 0,1). \label{2mstep3}
\end{equation}

The integrated form-factor 
\begin{equation}
\mathrm{Q}(x^{\|} -y^{\|},m,m_5) := \int_0^{\infty}d x^1 
	\int_0^{\infty} dy^1\,\mathrm{Q}_{\mathrm{4}}(x^{\|} -y^{\|},x^1,y^1,m,m_5)
\label{defQ4d}
\end{equation}
will play an important role here and in the subsequent section. In Appendix \ref{sec:some}, Eqs. \ref{QQ41}--\ref{QQ42}, we derive the following formula
\begin{equation}
\mathrm{Q}(x^{\|} -y^{\|},m,m_5) = -\frac{e^2 m_5}{m}\cdot G_0^{\mathbf{3D}}(x^{\|}-y^{\|},0)\cdot \Gh (x -\bar{y},2m,2m_5)\big|_{x^1 = 0 = y^1}
 \label{Q4d}
\end{equation}
where $G_0^{\mathbf{3D}}(x^{\|}-y^{\|},0)$ stands for the (massless) 3-dimensional Green's function defined in \eqref{G03d}.

Since $\Gh$ is proportional to $m$, see Eq.\ (\ref{defGh}), the limit $m\to 0$ in (\ref{Q4d}) can be done without any problem. By using this formula together with Eq.\ (\ref{largemlim}) we compute
\begin{equation}
\mathcal{C}=-\frac{\ii e^2}{2} \int d^3v^\| \, G_0^{\mathbf{3D}}(v^{\|},0)\cdot G_0^{\mathbf{3D}}(v^{\|},2)=
\frac {e^2}{16\pi}.\label{Codd2mResult} 
\end{equation}
Thus, performing the Pauli-Villars subtraction and going to the physical limit $m_{5,i}\to \infty$,
amount to adding Eq. \ref{factorized} to  \eqref{2mSodd} with overall weight equal to $c_1+c_2=-1$, and $\mathcal{C}$ given by the above expression. In this way  we obtain the renormalized parity odd effective action
\begin{equation}
S_{\mathrm{odd}}^{\mathrm{ren}}=\frac{\ii}{2} \int d^4x\, d^4y \, \varepsilon^{1ijk}\, \mathrm{Q}_4(x,y) A_i(x) \partial_{j[y]} A_k(y) + \frac{e^2\mathrm{sgn}(m_5)}{16\pi} \int d^3x^\| \varepsilon^{1ijk}A_i(x^\|,0)\partial_j A_k(x^\|,0).
\label{2mFINAL}
\end{equation}
We see, that the subtracted term is nothing else but the Chern-Simons action on the boundary with the level $k=\pm 1/4$. The action (\ref{2mFINAL}) vanishes in the limit $m_5\to\infty$ since the axial masses $m_{5,i}$ of regulator fields were taken of the same sign as $m_5$ (see Eq. \ref{factorized}). 

%%%%%
\section{Hall conductivity near the boundary}\label{sec:Hall}
In this section we compare the parity odd effective action (\ref{2mFINAL}) to its three-dimensional counterpart (\ref{Sodd3dRen}). The action (\ref{2mFINAL}) includes integration over the whole space, but the form-factor decays rapidly away of the boundary. To compare two actions, we propose to integrate the form-factors over the normal coordinates $x^1$ and $y^1$. Technically, this corresponds to plugging in (\ref{2mFINAL}) an electromagnetic potential that does not depend on the normal coordinate. Physically, we put the system in an external electromagnetic potential parallel to the boundary and constant in $x^1$ and measure the total current integrated over $x^1$. Due to the presence of $\varepsilon^{1ijk}$, the electric field in some direction parallel to the boundary leads to the current in a perpendicular direction (also along the boundary). Thus, we are dealing with a Hall type conductivity. The corresponding scalar form factor reads:
\begin{equation}
\mathrm{Q}(x^\| - y^\|)+\frac{e^2 \mathrm{sgn}(m_5)}{8\pi \ii } \delta(x^\|-y^\|). \label{Hall4D}
\end{equation}
This has to be compare with the corresponding form-factor for a $(2+1)$-dimensional Dirac fermion, that is given by
\begin{equation}
\mathrm{Q}_3(x^\| - y^\|)+\frac{e^2 \mathrm{sgn}(m_5)}{4\pi \ii } \delta(x^\|-y^\|), \label{Hall3D}
\end{equation}
see Eq.\ (\ref{Sodd3dRen}). We identified the coordinates in $3$D with coordinates on the boundary of the $4$D case, and the $3$D mass with $m_5$, as is suggested by the Dirac equation (\ref{bDir}) for boundary modes. We also took into account a sign factor in the Levi-Civita tensor, $\varepsilon^{1ijk}=-\varepsilon^{ijk}$. 

The relative strength of the effect in these two models is measured for $x^\|\neq y^\|$ by the fraction $\mathrm{Q}/\mathrm{Q}_3$. Equations\ (\ref{Q4d}) and (\ref{largemlim}) allow to derive the following relation
\begin{equation}
\mathrm{Q}(x^\|-y^\|,m,m_5)+\mathrm{Q}(x^\|-y^\|,-m,m_5)=\mathrm{Q}_3(x^\|-y^\|,m_5), \label{refl}
\end{equation} 
that permits to consider positive or negative masses only.

\begin{figure}[t]
\epsfxsize=2.8 in
\bigskip
\centerline{\epsffile{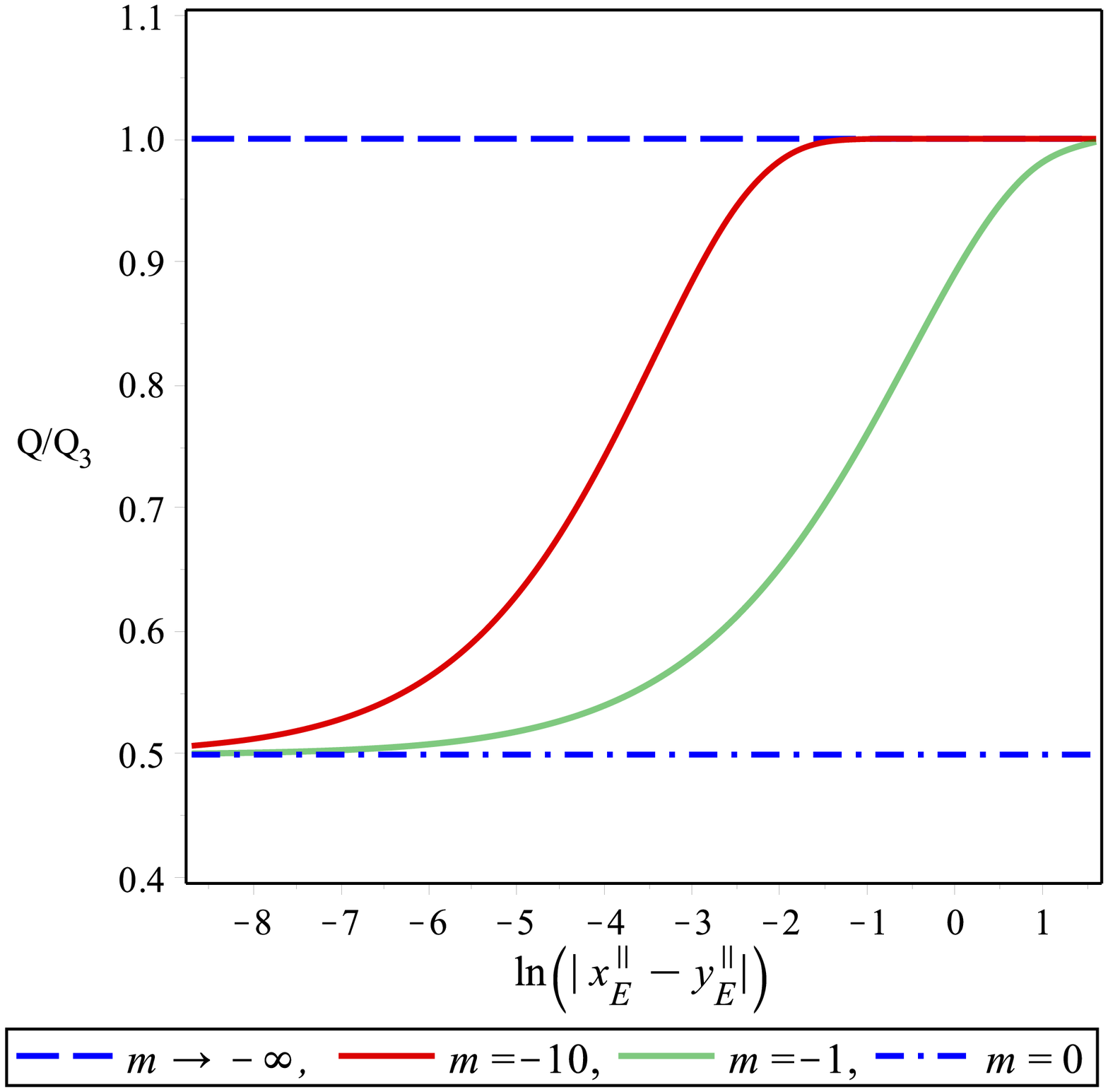}\epsfxsize=2.8 in\epsffile{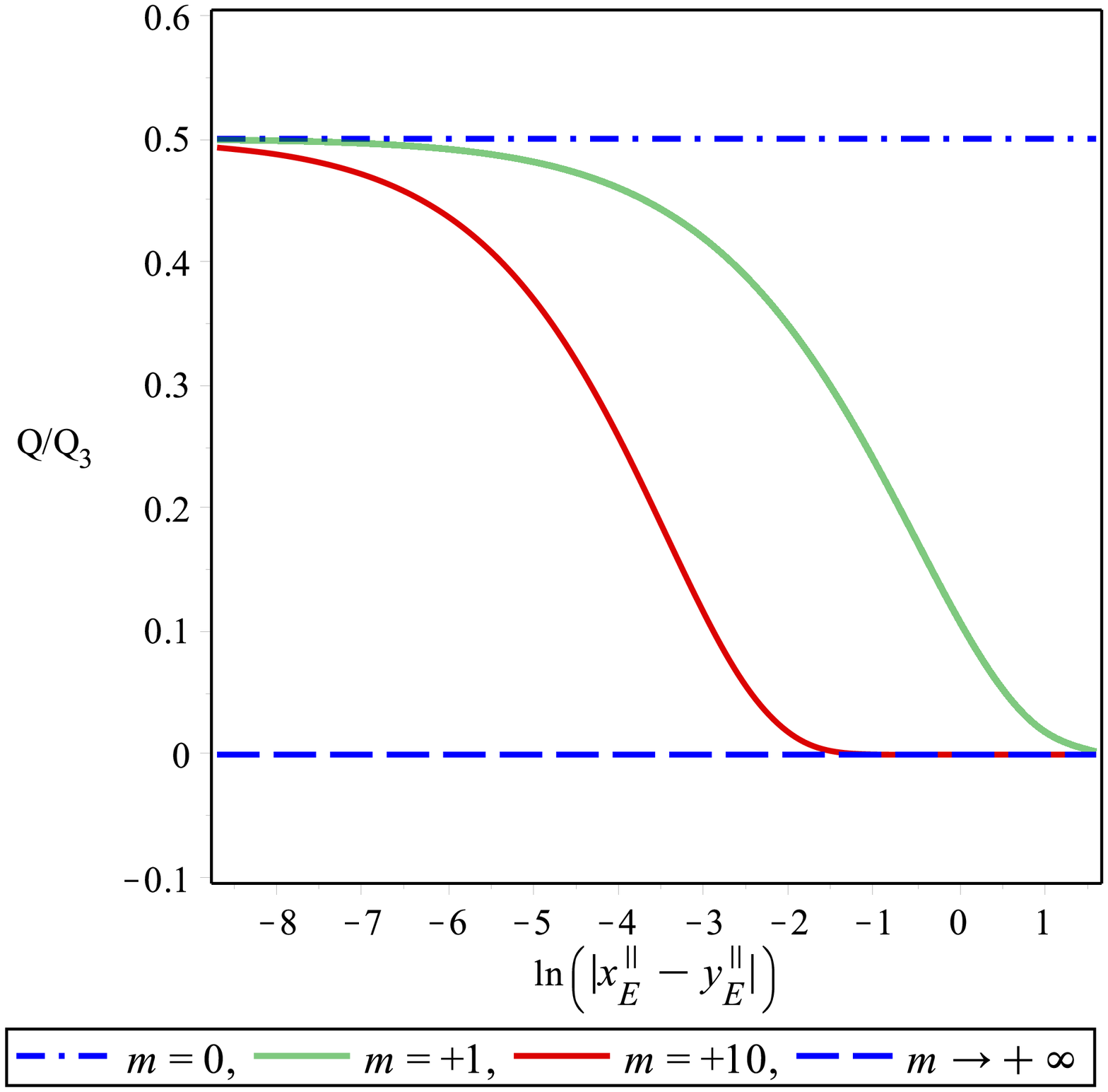}}
\caption{\sl The ratio of the integrated 4-dimensional form factor
$\mathrm{Q}(x^{\|} -y^{\|},m,m_5)$ and  the 3-dimensional form factor  $\mathrm{Q}_3(x^{\|} -y^{\|},m_5)$ at $m_5 = 1$. Blue lines correspond to the exact results \eqref{Qlimits1} and \eqref{Qlimits2}, whilst red and green curves are obtained numerically.
 \label{pic}}
\end{figure}

Some limiting cases may be studied analytically. In particular, in the small $m$ and short Euclidean distance limits we have
\begin{equation}
\lim_{|x^{\|}_E -y^{\|}_E|\to 0} \frac{\mathrm{Q}(x^{\|} -y^{\|},m,m_5)}{\mathrm{Q}_3(x^{\|} -y^{\|},m_5)} = \frac{1}{2} = \lim_{ |m|\to 0} \frac{\mathrm{Q}(x^{\|} -y^{\|},m,m_5)}{\mathrm{Q_3}(x^{\|} -y^{\|},m_5)}.
\label{Qlimits1}
\end{equation}
This $1/2$ combines nicely with the relative factor in front of delta function meaning a universal relative factor of $1/2$ for the polarization tensors at short distances.
In the opposite limit of large $|m|$ and large Euclidean distances the form factors behave as
\begin{eqnarray}
&&\lim_{|x^{\|}_E -y^{\|}_E|\to \infty} \frac{\mathrm{Q}(x^{\|} -y^{\|},m,m_5)}{\mathrm{Q_3}(x^{\|} -y^{\|},m_5)} = 0 = \lim_{|m|\to \infty} \frac{\mathrm{Q}(x^{\|} -y^{\|},m,m_5)}{\mathrm{Q_3}(x^{\|} -y^{\|},m_5)}\quad\mbox{for $m >0 $},\nonumber\\
&&\lim_{|x^{\|}_E -y^{\|}_E|\to \infty} \frac{\mathrm{Q}(x^{\|} -y^{\|},m,m_5)}{\mathrm{Q_3}(x^{\|} -y^{\|},m_5)} = 1 = \lim_{|m|\to \infty} \frac{\mathrm{Q}(x^{\|} -y^{\|},m,m_5)}{\mathrm{Q_3}(x^{\|} -y^{\|},m_5)} \quad\mbox{for $m < 0 $}. \label{Qlimits2}
\end{eqnarray}
The second line in the equation above follows from the first one by Eq.\ (\ref{refl}). For non-asymptotic values of the parameters the fraction of form factors is depicted at Fig.\ \ref{pic}. We use Wick-rotated coordinates and Euclidean distance to simplify the problem. The Euclidean regime is sufficient to describe some quantum phenomena, like the Casimir effect, though it does not tell us much about the optical properties of Topological Insulators.

Since typical topological insulators have massless boundary states, the most important limit is $m_5\to 0$. As follows from the explicit expressions (\ref{QQ42}) and (\ref{Q3d}),
\begin{equation}
\lim_{m_5\to 0} \mathrm{Q}(x^\|-y^\|,m,m_5)=\lim_{m_5\to 0} \mathrm{Q}_3(x^\|-y^\|,m_5)=0.\label{QQm5}
\end{equation}
Thus, in this limit, the parity odd parts of both effective actions are given by local Chern-Simons terms. The Chern-Simons level of effective boundary theory $k=\mathrm{sgn}(m_5)/4$ does not depend on the bulk mass $m$ and is exactly one half of the Chern-Simons level for a Dirac fermion in $(2+1)$ dimensions. In other words, the Hall conductivity on the boundary of a topological insulator without surface gap is one half of that for a single massless Dirac field. For $m=0$, this result was established in \cite{Kurkov:2017cdz}, where the physical meaning of this apparently surprising relation was discussed in detail.

We recall, that the dependence of form factors on the Fermi velocity is restored by the rescalings (\ref{vF1}). The limits (\ref{Qlimits1}), (\ref{Qlimits2}) and (\ref{QQm5}) remain valid. The curves on Fig.\ \ref{pic} were drawn for $m_5=1$ and may be interpreted also in terms of dimensionless variables $m/m_5$, $|(x^\|_E-y^\|_E) m_5|$.  The fraction of two masses remain unchanged under the rescaling, as also $|(x^\|_E-y^\|_E) m_5|$ does if the separation $(x^\|_E-y^\|)$ is in the (Euclidean) time directions. For spatial separations, the rescaling leads to $|(x^\|_E-y^\|_E) m_5|\to |(x^\|_E-y^\|_E) m_5|v_F^{-1}$. For reasonable values of the parameters, $m=0.1{\rm eV}$, $m_5=0.01{\rm eV}$, $v_F=10^{-3}$, the fraction $Q/Q_3$ assumes its asymptotic values $1$ and $0$ for spatial separations larger than a few Angstr{\"o}m. This is a position space counterpart of the phenomenon that we have already discussed at the end of Sec.\ \ref{sec:nb}: the dependence of renormalized physical quantities on spatial momentum is much weaker in this model than the dependence of the same quantities on the frequency.

%%%%
\section{Conclusions}\label{sec:con}
Here we give a short summary of the main results obtained in this work. We suggested a modification of the PV regularization scheme that consists in giving axial masses to the PV regulators that become infinite in the physical limit. We demonstrated, that this scheme indeed renormalizes the polarization tensor of Dirac fermions in half-space, though the usual PV prescription fails to produce finite results. We computed the renormalized polarization tensor, that appeared to be given by a sum of rather complicated expressions. However, after the integration over the normal coordinate the parity odd part of polarization tensor became relatively simple and admitted a comparison to the corresponding quantity for a $(2+1)$ dimensional Dirac fermion. This part (corresponding to a distance-depending Hall type conductivity) was analyzed in detail. Our results have some quite immediate applications, like e.g. to the study of possibility of the Casimir repulsion between topological insulators (see \cite{Fialkovsky:2018fpo} and references therein), that we are planning to address in the future.

\begin{acknowledgments}
One of us (D.V.) is grateful to Juan Mateos Guilarte for fruitful conversations. 
This work was supported in parts by the S\~ao Paulo Research Foundation (FAPESP), projects 2016/03319-6 and 2017/50294-1 (SPRINT), by the grants 303807/2016-4 and 428951/2018-0 of CNPq, by the RFBR project 18-02-00149-a and by the Tomsk State University Competitiveness Improvement Program.  M.K. acknowledges the support of the INFN Iniziativa Specifica GeoSymQFT.
\end{acknowledgments}

\appendix
%%%%
\section{Polarization diagram in 3 dimensions}\label{sec:3D}
The polarization tensor of Dirac fermions in $(2+1)$ dimensions was computed in the momentum representation long ago \cite{Appelquist:1986fd,Niemi:1983rq,Redlich:1983dv}, see also \cite{Fialkovsky:2011wh}. The mass dependence of parity anomaly in three dimensions was studied in detail in \cite{Deser:1997gp}.
In this Appendix we rederive the parity odd part of effective action for photons in a $3$D flat space without boundaries in the coordinate representation. Since the rule for recovering the $v_F$ dependence of polarization tensor in $3$D is the same as in $4$D modulo a restriction on the range of the indices, see (\ref{vF12}), we make the computations for $v_F=1$. We consider a single Dirac fermion in $2+1$ dimensions with free Dirac operator
\begin{equation}
\slashed{D}_{0\,\mathbf{3D}} = \ii\Gamma^j\partial_{j} + m,
\end{equation}
with the gamma matrices $\Gamma^j$, $j=0,2,3$, satisfying usual Clifford algebra relations and $\mathrm{tr}\, \bigl(\Gamma^i\Gamma^j\Gamma^k\bigr)=-2\ii \varepsilon^{ijk}$, $\varepsilon^{023}=1$, cf. Eq.\ (\ref{bDir}). Similarly to the 4D case,
\begin{equation}
\slashed{D}_{0\,\mathbf{3D}}^{-1}(x,y) = \overline{\slashed{D}}_{0\,\mathbf{3D}}(x) G_0^{\mathbf{3D}}(x-y)
\end{equation}
with
\begin{equation}
\overline{\sD}_{0\,\mathbf{3D}}=\ii\Gamma^j\partial_{j} - m,\qquad
G_0^{\mathbf{3D}}(x-y) =  -\frac{\ii}{4\pi}\frac{e^{-|m| \,|x-y|}}{|x-y|}.  \label{G03d}
\end{equation}
As in the 4D case, $|x-y|= \sqrt{{-(x-y)^i(x-y)^jg_{ij} + \ii 0}}$ and $ g=\mathrm{diag}\, (+1,-1,-1)$.

The quadratic part of one-loop effective action for electromagentic field reads
\begin{equation}
S_{\rm eff}[A,m] 
	=  \frac{\ii e^2}{2} \,  \int d^3x\, d^3y\,  \,
\mathrm{tr}\left(\Gamma^i\slashed{D}_{0\,\mathbf{3D}}^{-1}(x-y)\Gamma^j\slashed{D}_{0\,\mathbf{3D}}^{-1}(y-x)\right)   
A_{i}(x)A_{j}(y)\,
\label{SeffCoord3d}
\end{equation}
with its' parity odd part (containing an odd number of gamma matrices under the trace) being
\begin{equation}
S_{\mathrm{odd}} = -\ii m e^2\int d^3 x  d^3 y\,
\varepsilon^{ijk} \Big(G_0^{\mathbf{3D}}(x-y)\Big)^2
A_i(x) \partial_{j}A_{k}(y).
\label{Sodd3d}
\end{equation}

It is very well known, that to make the action (\ref{SeffCoord3d}) finite, a single PV subtraction is sufficient. Note, that though the odd part (\ref{Sodd3d}) is finite, the subtraction has to be done in this part as well. After subtracting from (\ref{Sodd3d}) the contribution of a spinor with mass $M$ of the same sign as $m$ and taking the limit $|M|\to \infty$, we obtain
\begin{equation}
S_{\mathrm{odd}}^{\mathrm{ren}} = -\frac{\ii}2 \int d^3 x  d^3 y\,\,
	\varepsilon^{ijk} \Big(Q_3 (x-y)+ \frac{{e^2 \mathrm{sgn}(m)}}{4\pi\ii} \delta(x - y)\Big)
	A_i(x) \partial_{j}A_{k}(y),
	\label{Sodd3dRen}
\end{equation}
where
\begin{equation}
\mathrm{Q}_{3}(x -y ) = 2e^2 m\,\,G_0^{\mathbf{3D}}(x-y)^2. \label{Q3d}
\end{equation}
The basic property of renormalized action (\ref{Sodd3dRen}) is that it vanishes in the limit $|m|\to \infty$.

%%%%
\section{Some useful formulas involving Green's functions}\label{sec:some}
Let $x_E$ be the Wick rotated coordinate, $x^0 \to -\ii x^4$. The following asymptotic expansions at $|x_E|\to 0$ can be checked by using the explicit form of propagator (\ref{scprop}):
\begin{eqnarray}
&&G_0(x_E)\simeq -\frac{\ii}{4\pi^2}\frac{1}{x^2_E} -\frac{\ii}{8\pi^2}(m^2 + m_5^2) \ln (|x_E|) +\mathcal{O}\Big( |x_E|^0 \Big),
\nonumber\\
&&\partial_\mu G_0(x_E) \simeq (\partial_\mu |x_E|) \cdot\left(\frac{\ii}{2\pi^2}\frac{1}{|x_E|^3}  -\frac{\ii}{8\pi^2}(m^2 + m_5^2) \,\frac{1}{|x_E|} +\mathcal{O}\left(|x_E|\ln \big( |x_E|\big)\right) \right).
\label{G0SingStruct}
\end{eqnarray}
The singularities of $\Gh(x)$ also appear at $|x_E|=0$, but they depend on the angle $\phi$,
\begin{equation}
\tan(\phi) = \frac{x^1}{|x^\| |}, \quad 0 \leq \phi \leq \frac{\pi}{2}. \label{angledef}
\end{equation}
After lengthy but otherwise straightforward computations we obtain the following estimates
\begin{eqnarray}
&&\Gh (x_E)\simeq - \frac{\ii m}{4\pi^2}\cdot\frac{\phi - \frac{\pi}{2}}{\cos(\phi)}\cdot \frac{1}{|x_E|}
+ \frac{\ii m^2}{4\pi^2} \ln (|x_E|) + {\mathcal{O}(|x_E|^0)}, \nonumber\\
&&\partial_{i}\Gh(x_E)\simeq
\frac{\partial |x^\| |}{\partial x^{i}}\Bigg( -\frac{\ii\,m}{8\pi^2}\cdot\frac{ - 2\phi + \pi - \sin(2\phi)}{\cos^2(\phi)}\cdot \frac{1}{|x_E|^2} 
- \frac{\ii m^2}{8\pi^2} \cdot\frac{ (- 2\phi + \pi)\sin(\phi) - 2\cos(\phi)}{\cos^2(\phi)}\cdot\frac{1}{|x_E|} 
 +  {\mathcal{O}(|x_E|^0)} 
\Bigg), \nonumber\\
&&\partial_1\Gh(x_E)\simeq -\frac{\ii m}{4\pi^2}\frac{1}{|x_E|^2} - \frac{\ii m^2}{4\pi^2}\cdot\frac{\phi - \frac{\pi}{2}}{\cos(\phi)}\cdot \frac{1}{|x_E|}
+
\frac{\ii m}{8\pi^2}(m^2 - m_5^2) \ln(|x_E|) +\mathcal{O}(|x_E|^0). \label{Gestim}
\end{eqnarray}
The correction terms $\mathcal{O}(|x_E|^0)$  {are uniformly bounded over $\phi\in [0;\pi/2]$}. Note, that all angular functions in (\ref{Gestim}) are non-singular.

In the Euclidean region the Green's function $G_0$ admits a proper time integral representation
\begin{equation}
G_0(x_E) =  -\frac{\ii}{16\pi^2} \int_0^\infty \frac{d t}{t^2}\, \exp{\left(-\frac{|x_E|^2}{4t} - \mfm^2 \,t\right)}.
\label{proptimerepress}
\end{equation}
This equation can be integrated over $x^1$ with weight $e^{-x^1m}$ yielding
\begin{equation}
\int_{0}^{\infty} dx^1\, e^{-x^1m} G_0(x_E)=- \frac{\ii}{16\pi^{3/2}} \int_0^\infty \frac{d t}{t^{3/2}}\, \exp \left(-\frac{|x^{\|}_E|^2}{4t} - m_5^2\,t \right) \,\mathrm{erfc}\big(m \sqrt{t}\big).  \label{appfcalc}
\end{equation}
By taking into account the proper time representation for $3$D propagator
\begin{equation}
G_0^{\mathbf{3D}}(x^{\|}_E;m_5)=- \frac{\ii}{8\pi^{3/2}} \int_0^\infty \frac{d t}{t^{3/2}}\, \exp \left(-\frac{|x^{\|}_E|^2}{4t} - m_5^2\,t \right),
\end{equation}
combining (\ref{appfcalc}) for $m$ and $-m$, and rotating back to the Minkowski signature, one gets
\begin{equation}
\int_0^\infty dx^1\left( e^{-x^1m}+e^{x^1m}\right)G_0(x)=G_0^{\mathbf{3D}}(x^{\|};m_5).\label{largemlim}
\end{equation}

Next, we derive Eq.\ (\ref{Q4d}) for the integrated formfactor $\mathrm{Q}$. Through a sequence of manipulations with integrals, which includes changes of variables and integrations by parts, one arrives at
\begin{equation}
{\mathrm{Q}}=4m_5e^2 \int_0^{\infty} d x^1 \int_{-\infty}^{+\infty} d y^1 \,e^{-2 x^1 m}\,
   {{G}_0\big(x-y\big)}\,  {G}_0\big(x-\bar{y}\big).\label{QQ41}
\end{equation}
In this formula, we perform the Wick rotation, use the proper time representation of the Green's functions and integrate over $x^1$ and $y^1$ to obtain
\begin{equation*}
{\mathrm{Q}}=-\frac{m_5e^2}{64 \pi^3}\int_0^{\infty} d t \int_{0}^{\infty} d \tau \,\frac{
  1}{t^{\frac{3}{2}}\tau^{\frac{3}{2}}} \cdot\exp{\left(-(t+\tau)m_5^2 - \left(\frac{1}{t} + \frac{1}{\tau}\right)\frac{|x^{\|}_E-y^{\|}_E|^2}{4}\right)}
\cdot\mathrm{erfc}{\left(m\sqrt{t+\tau}\right)} . 
\end{equation*}
After change of the variables,
\begin{equation*}
t = r \cos^2(\phi), \quad
\tau = r \sin^2(\phi), \quad
r\in [0;\infty),\quad \phi \in[0;\pi/2],
\end{equation*}
the integration over $\phi$ is easily performed yielding
 \begin{equation*}
{\mathrm{Q}}=-\frac{m_5e^2}{16 \pi^{\frac{5}{2}}}\cdot \frac{1}{|x^{\|}_E - y^{\|}_E|}\cdot \int_0^{\infty}
  \frac{d r}{r^{\frac{3}{2}}}\, \exp{\left(-\frac{|x^{\|}_E - y^{\|}_E|^2}{r} - r\cdot m_5^2\right)}\cdot\mathrm{erfc}{\left(m\,\sqrt{r}\right)}. 
\end{equation*}
After changing $r=4t$, using the relation \eqref{appfcalc} and continuing the result to Minkowski space, one gets
\begin{equation}
\mathrm{Q} = - \frac{e^2m_5}{m}\cdot \frac{-\ii}{4\pi |x^{\|} - y^{\|}|}  \cdot (-2m) \int_{0}^{\infty} d z^1 \,e^{-z^1\,( 2m) }\,G_0(z , 2\mathfrak{m})\big|_{z^{\|} = x^{\|}-y^{\|}}
,\label{QQ42}
\end{equation}
where one recognizes Eq.\ (\ref{Q4d}).

\bibliography{graphene,parity}

\end{document}